\documentstyle{article}
\begin{document}

\title{Lattice  Boltzmann-Langevin Equations\footnote{To
appear in
{\em Pattern formation and Lattice Gas Automata}
(Proceedings Nato Workshop,  Waterloo, Canada, June 7-12,
1993), A. Lawniczak and R. Kapral, eds,
{\em Fields Institute Communications} ( American Math.
Society)}}
\author{J.W. Dufty\thanks{Department of Physics, University
of Florida,
Gainesville, FL32611, USA} \,and M.H. Ernst\thanks{Institute
for  Theoretical Physics, University of Utrecht, The
Netherlands}} \maketitle

\begin{abstract}
Intrinsic fluctuations around the solution of the lattice
Boltzmann equation are described or modeled by addition of a
white Gaussian noise source. For stationary states a
fluctuation-dissipation theorem relates the variance of the
fluctuations to the linearized Boltzmann collision operator
and the pair correlation function.
\end{abstract}

\section{Introduction}
\indent The discussions and contributions at  this meeting
have shown a
growing interest in fluctuations both in the microscopic
lattice gas approach, as well as in the phenomenological
lattice Boltzmann approach. The emphasis was on unstable
systems (phase separation, spinodal decomposition, pattern
formation) where fluctuations drive the system away from a
spatially uniform equilibrium state.  So far the noise
considered in connection with the lattice Boltzmann approach
was introduced in a rather ad hoc fashion.  It may be
additive or multiplicative,  external or intrinsic.  There
do not seem to be any systematic studies that make the
connection with the multitude of results for fluctuations
and transport in the continuous case (see  Bixon and Zwanzig
[1969], Fox and Uhlenbeck [1970], Ernst and  Cohen [1981]
and Marchetti and Dufty [1983]). The goal of this paper is
to provide this link and to show that the well-etablished
results on intrinsic fluctuations, derived from the
Boltzmann-Langevin equation in the continuous case, can be
extended to the lattice Boltzmann equation.

\indent The Boltzmann  equation is a deterministic equation
for the
average occupation $f(\vec r,\vec c,t)$ of a one particle
state specified by position $\vec r$ and velocity $\vec c$.
When derived from an underlying microdynamics, it describes
the `slow' time evolution of $f(\vec r,\vec c,t)$ on coarse
grained spatial and temporal scales of the order of the mean
free path $\ell_o$ and time $t_o$ between collisions. The
deviations of the mesoscopic occupation, $n (\vec r,\vec
c,t)$, from this average Boltzmann value reflect
fluctuations due to other `fast' degrees of freedom that are
averaged out in obtaining the Boltzmann equation. A measure
of these fluctuations is given by their correlation
functions, which also can be calculated from the
microdynamics. A unified picture of both the transport and
fluctuations is given by a Boltzmann-Langevin (B-L)
equation.  The existence of other degrees of freedom is
recognized in the B-L equation by adding a stochastic source
that generates fluctuations in the solutions to this
equation relative to the deterministic solution. This is a
particularly convenient representation for simulation of
noise effects on the Boltzmann equation. The characteristics
of this stochastic source should be determined from the
microdynamics for consistency with the direct evaluation of
correlation functions using methods of statistical mechanics
and kinetic theory. Studies along these lines are in
progress [Dufty and Ernst 1994].

\indent In other contexts, a Boltzmann equation is simply
postulated as
a mathematical model to simulate and study macroscopic fluid
properties. Its extension to a corresponding B-L equation to
describe the effects of noise now has no underlying
microdynamics to characterize that noise. To assist in this
characterization, we establish in the next section the
general relationship of the noise variance to the pair
correlation function at equal time and different times.  The
form of this relationship is independent of any underlying
microdynamics. This connection of the noise to correlation
functions  provides a somewhat more physical basis for
suggesting the noise characteristics. Furthermore, known
results about the correlation functions  for systems with a
microdynamics can be `borrowed' for application to the
mathematical modeling as well.

\indent Our primary interest here is to make the connections
just mentioned and to discuss their content at a
phenomenological level. The B-L equation is defined in the
next section and it is assumed that the noise is white
(uncorrelated in time) and essentially characterized by its
variance (this is strictly the case only for Gaussian
noise). Equations for the correlation of fluctuations at
equal and different times are derived from the B-L equation,
under the condition that the fluctuations are small. It is
observed that these equations are the precise analog of
those derived for a real low density gas (see Ernst and
Cohen [1981], Marchetti and Dufty [1983]).
 For the special case of fluctuations about a stationary
state a fluctuation-dissipation relation is obtained,
defining the noise variance in terms of the stationary state
correlations and the linear Boltzmann operator. For
non-stationary states additional information is required.
The analysis applies as well to unstable states up to times
for which the fluctuations remain small. In section 2 we
give a concrete example of a BGK-Langevin description of
fluctuations about a steady state. Finally, in section 3 we
show that hydrodynamic equations with noise can be obtained
consistently from the B-L description. The expected
Landau-Lifshitz form is confirmed.

\section{B-L Equation and Correlation Functions}

\indent In the  absence  of  noise,  the  lattice  Boltzmann
 equation
describes deterministic dynamics  for  the  occupation
number, $n_i(\vec r,t)$,  of  a single particle state at
node $\vec r$ in velocity channel $\vec{c}_{i}$ where $i$
labels the allowed velocity states.  The $n_i (\vec r,t)$
are continuous real variables on  the interval $[0,a]$
where  $a$  is  the  maximum occupancy of a state (in the
case of an underlying  microdynamics  with  Fermi exclusion
rules, a value $a > 1$ implies that these occupation numbers
are  coarse grained averages over  several  nodes). The
corresponding B-L equation with noise is,
\begin{equation} \label{e1}
 {n}(x,t+1) = V(x\mid n(t)) +  S_xF(x,t),
\end{equation}
\noindent where a simplified notation $n(x,t)$ = $
{n}_i(\vec r,t)$ has  been
introduced.  The generator for the  deterministic  dynamics,
$ V(x\mid n(t))$, is given by,
\begin{equation} \label{e2}
V (x\mid n(t)) = S_x(n(x,t)+ I(x\mid n(t))),
\end{equation}
\noindent where $I(x\mid n(t))$ = $I_i(\vec  r  \mid  n(t))$
 is  the
non-linear Boltzmann collision operator,  and $S_x$ is the
free streaming operator defined by $S_x f(\vec r) = f(\vec r
- {\vec c}_i)$ for any  function of $\vec r$. Finally, the
stochastic source, $ F (x,t)$, is assumed to  represent
Gaussian  white  noise  with  zero  mean  value,
statistically independent of the occupation numbers,
$\langle n (x,t) F (y,\tau)\rangle = 0$ for $\tau \leq t$,
and covariance  specified  by  a  noise  intensity  matrix,
$B(x,y)$,
\begin{equation} \label{a3}
\langle F(x,t)F(y,t^\prime) \rangle = \delta(t,t^\prime)
B(x,y\mid n(t)).
\end{equation}
The average is taken over the noise source.  Furthermore the
noise does not generate any mass, momentum or energy, so
that $F$ is orthogonal to the summational invariants.  We
note that the noise matrix $B(x,y\mid n)$ in general can
depend on the state at time t; this is an example of
multiplicative noise. Once the noise matrix has been
specified, the above provides a complete description of
fluctuations and transport.

\indent  To simplify the notation, let {\bf n} denote the
vector whose
components are $n(x,t)$. Similarly, {\bf V}({\bf n}), {\bf
F}, and {\bf B(\bf n)}  denote the vectors and matrix with
elements $V(x |{n})$, $F(x)$, and $B(x,y\mid n)$,
respectively. The B-L equation is then,
\begin{equation} \label{j1}
{\bf n}(t+1) =  {\bf V}({\bf n}(t)) + {\bf S}{\bf F}(t), \>
\> \> \langle {\bf F}(t)
{\bf F}(t^\prime)\rangle
= \delta (t,t^\prime){\bf B}({\bf n}(t)).
\end{equation}
The solution to the deterministic (noise-free) Boltzmann
equation, ${\bf f}(t)$, is obtained from
\begin{equation} \label{j2}
{\bf f}(t+1) =  {\bf V}({\bf f}(t)).
\end{equation}
The fluctuations around this deterministic solution are
measured by $\delta {\bf n}(t)$ = $ {\bf n}(t) - {\bf
f}(t)$. The corresponding pair correlation functions are
defined by,
\begin{equation} \label{j3}
{\bf g}(t)  = \langle\delta {\bf n}(t)\delta {\bf
n}(t)\rangle, \> \> \>
{\bf c}(t,t^\prime)  = \langle\delta {\bf n}(t)\delta {\bf
n}(t^\prime)\rangle.
\end{equation}
Closed equations for these correlation functions follow in
two steps: first the B-L equation is applied to $ {\bf
g}(t+1)$ and ${\bf c}(t+1,t^\prime)$; next the resulting
averages are expanded to lowest order in $\delta {\bf
n}(t)$, for small fluctuations,
\begin{equation} \label{j5}
 {\bf g}(t+1) = {\bf L}(t) {\bf g}(t){\bf L}^T (t) + {\bf
S}{\bf B}(t){\bf S}^T
\end{equation}
\begin{equation} \label{j6}
 {\bf c}(t+1,t^\prime) = {\bf L}(t) {\bf c}(t,t^\prime).
\end{equation}
Here $\bf {B}(t) \equiv \bf {B}({\bf f}(t))$ and ${\bf
L}(t)$ is the matrix,
\begin{equation} \label{j7}
{\bf L}(t) \equiv {d\over d{\bf n}}{\bf V (n)}\mid_{{\bf
f}(t)}.
\end{equation}
Both the equal time and two time correlation functions obey
linear equations governed by the matrix {\bf L}(t) whose
form is determined from the generator, ${\bf V}({\bf f})$,
for the given Boltzmann equation. They depend on the
specific nonequilibrium solution to the Boltzmann equation,
{\bf f}(t), being considered. For the special case where
${\bf f}(t) = {\bf f}_o$ is the equilibrium state, this
result forms the starting point in the calculation of the
Green-Kubo formulas for lattice gas cellular automata
(LGCA) in the mean field approximation [Brito et al 1991],
as well as for the dynamic structure factor in such systems
[Grosfils et al 1993].

\indent We remark at this point that (\ref{j2}) and
(\ref{j5}) through
(\ref{j7}) are the direct analog of corresponding results
from the kinetic theory of real gases at low density. The
difference here is that the form of ${\bf B}(t)$ is given
explicitly from the theory there, whereas here it is as yet
unspecified.

\indent Before discussing the content of these equations it
may be
useful to see how they can be applied to determine the noise
matrix in a special case. Consider fluctuations around a
stationary (or quasi-stationary) solution to the Boltzmann
equation, ${\bf f}_o$. For consistency, the noise matrix
must be chosen such that the equal time correlation
functions are also stationary. Consequently, (\ref{j5})
gives,
\begin{equation} \label{j8}
{\bf S}{\bf B}_o{\bf S}^T =
{\bf g} - {\bf L}_o {\bf g}{\bf L}_o^T \> \qquad {\rm or}
\qquad
{\bf B}_o = {\bf S}^T {\bf g}{\bf S} -  (1+{\bf {\Omega}})
{\bf g}(1+{\bf {\Omega}}^T),
\end{equation}
where ${\bf B}_o$ and ${\bf L}_o$ are given by $\bf {B}$ and
${\bf L}$  evaluated at the stationary state ${\bf f}_o$,
and ${\bf L}_o$ = ${\bf S}(1+{\bf{\Omega}})$ has been
expressed in terms of the linearized collision operator
${\bf{\Omega}}$. This is the classical
fluctuation-dissipation theorem extended to lattice gas
automata. It expresses the noise matrix in terms of the
stationary state correlations and the Boltzmann collision
operator linearized around the stationary state. A more
explicit form is given in the example at the end of this
section.

\indent How would (\ref{j8}) be used to implement a
simulation of noise in
a B-L equation?  First the model for the Boltzmann equation
is specified by using  collision rules for a LGCA or by a
BGK model; next the equal time correlation matrix is
determined for the model; finally, solutions to the B-L
equation are simulated for initial states near ${\bf f}_o$,
selecting the noise ${\bf F}$ from a Gaussian whose half
width ${\bf B}_o$ is determined from the
fluctuation-dissipation relation. In this simulation of the
B-L equation in the stationary state the equal time
correlation function $\bf g$ is used as input to determine
the noise statistics ${\bf B}_o$, and the two time
correlation function ${\bf c}(t,t^\prime) = \overline
{\delta {\bf n}(t)\delta {\bf n}(t^\prime)}$ can be measured
by simulations, even for models without any underlying
microdynamics.  The simulated values should agree with the
theoretical value calculated from (\ref{j6}).

\indent To illustrate the procedure sketched below
(\ref{j8}) we consider
two examples: a standard LGCA that obeys the conditions of
semi-detailed balance (SDB), and a BGK-model. In the LGCA
all equilibrium fluctuations are totally uncorrelated, i.e.,
\begin{equation} \label{a3b}
g(x,y) =\delta(x,y) g_i   \qquad (x=\vec r\ i; y=\vec r\
^\prime j)
\end{equation}
with a known value, $g_i = f_{oi}(1-f_{oi})$ for the
fluctuations in a single particle state because of the Fermi
exclusion rule. Furthermore, the collision rules are
strictly local such that the collision matrix in (\ref{j8})
has the form $\Omega (x,y) =
\delta (\vec r, \vec r \,^\prime)\Omega_{ij}$.   The
noise strength (\ref{j8}) reduces to
\begin{equation} \label{a5}
 B(x,y) =\delta(\vec r, \vec r\,^\prime) B_{ij} =
 -\delta(\vec r, \vec r\,^\prime)  \left[\,\Omega_{ij} g_j +
\Omega_{ji} g_i
+ \Omega_{i\ell} g_\ell \Omega_{j \ell} \right].
\end{equation}
\indent Let us compare these results with
those of Bixon and Zwanzig [1969], and Fox and Uhlenbeck
[1970] for the strength of the fluctuating term in the
continuous case.  The first two terms are direct analogs of
the corresponding terms in the continuous case.  The term
quadratic in $\Omega$ is a typical lattice effect, similar
to the so called `propagation part' of the transport
coefficients, as occurring in lattice gas automata (see e.g.
Ernst and  Dufty [1991]) and in the lattice Boltzmann
approach. This will be shown in section 3.

\indent As a second example we consider a typical athermal
BGK-model
equation,
\begin{equation} \label{b1}
f_i(\vec{r}+\vec{c}_i,t+1) - f_i(\vec{r},t) =
- \frac{1}{\tau}\left[f_i(\vec{r},t)
 - f_i^{\ell}\right],
\end{equation}
\noindent where the local equilibrium distribution
$f_i^{\ell}$ depends
on the local density $\rho(\vec r,t)  = \Sigma
_{i}f_{i}(\vec r,t)$ and the local flow velocity $\rho(\vec
r,t) \vec{u}(\vec r,t) =
\Sigma _{i}\vec{c}_{i}f_{i}(\vec r,t)$.  We linearize  this
equation  around
basic equilibrium $f_{oi}$ with vanishing flow velocity,
where $f_{oi}$ equals  $f_{o}$ for  a rest particle $(i=0)$
and  equals $f$  for  a  moving  particle.
  The collision operator is diagonal and can be identified
as,
\begin{equation} \label{b2}
\Omega_{ii} = - \frac{1}{\tau}\left[1-
\frac{df_i^{\ell}}{d\rho} -
(\frac{\vec c_i}{\rho})\cdot
\frac{df_i^{\ell}}{d\vec u}\right]_o ,
\end{equation}
\noindent where the subscript (o) indicates that the
derivatives are taken at
 $\vec{u} = 0.$
\par
As a specific example we take the BGK-model defined in
Eq.(15)  of  Ernst and Bussemaker (see these proceedings)
and obtain,
\begin{equation} \label{b3}
\Omega_{oo} = - \frac{dc_s^2}{\tau c_o^2}, \qquad
\Omega_{ii} =
- \frac{1}{\tau}\left[1-
\frac{d}{bc_o^2}(c_s^2+c_o^2)\right],
\end{equation}
\noindent where i refers to a moving particle.  Furthermore,
$d$ is the
spatial dimension of the  lattice, $\mid \vec{c}_{i}\mid  =
c_{o}$  the lattice distance, and $b$ the coordination
number.  We also  used the speed of sound from the above
reference, i.e.,
\par
\begin{equation} \label{b4}
c_s^{2} = {dp\over d\rho } = {b{c_o}^{2}\over d} {df\over
d\rho } =
{c^{2}_{o}\over d} (1-{df_{o}\over d\rho }).
\end{equation}
\noindent So far we have specified the matrix elements of
the collision
operator to  be used in the expression (\ref{a5}) for the
noise strength. Next  we  address  the choice of equilibrium
fluctuation $g_{i} =
\langle(\delta {n}_{i})^{2}\rangle$.  In lattice gas
automata the exclusion  rule
for occupation numbers makes the  equilibrium  fluctuations
$g_{i}$ similar to those for an ideal Fermi gas. In the
mathematical modeling of the lattice Boltzmann approach the
positive function $g_{i}$ can be chosen  freely  by lack of
an underlying microscopic model.  However, the choice $g_{i}
= \rho (df^{o}_{i}/d\rho )$, which resembles the result for
a Maxwell-Boltzmann gas, is closest  in  spirit to the
BGK-model.  This is  so  because  the  local  equilibrium
distribution function $f^{\ell }_{i}$,  used  in  typical
lattice  gas  applications
[Chen et al 1992], depends on the local average density
$\rho $ and the fluid  flow  velocity $\vec{u}$,  and
resembles a local Maxwell-Boltzmann distribution expanded to
terms of ${\cal O}(u^{2})$ for small velocities.  In the
typical model under consideration we  have for  the rest
particles $g_{o} = \rho (1-dc^{2}_{s}/c^{2}_{o})$ and for
moving particles $g_{i} = (\rho d/b)(c^{2}_{s}/c^{2}_{o})$.
This completes the determination of the linear BGK-Langevin
equation.
\par

\section{Noise in Fluid Dynamics}

\indent In this section we calculate the correlation
function in
equilibrium of the fluctuating part of the single particle
distribution function  in the limit of large spatial and
temporal scales, for systems where the noise strength is
determined by $B_o(x,y)$ in (\ref{a5}). From that result one
can compute the fluctuations in the pressure tensor or in
the heat current, if energy is also conserved  in the
lattice Boltzmann equation. The results obtained will be
compared with the Landau-Lifshitz formulas for the
correlation strength in fluctuating hydrodynamics.

\noindent
For the analysis of this section it is convenient to work
with Fourier and Laplace transforms, defined through,
\begin{equation} \label{a12}
\tilde{n}_i(\vec k,z) = \sum^\infty_{t=0} \sum_{\vec r}
{\rm e}^{-i \vec k \cdot \vec r-zt} \delta {n}_i(\vec r,t) =
 \sum^\infty_{t=0} {\rm e}^{-zt} n_i(\vec k,t).
\end{equation}
For fluctuations around equilibrium the B-L equation
becomes,
\begin{equation} \label{a12a}
e^{i\vec k \cdot \vec c_i+z} \tilde{n}_i(\vec k,z) =
(\delta_{ij} +
\Omega_{ij})\tilde{n}_i(\vec k,z)  + {\tilde F}_i(\vec k,z).
\end{equation}
In vector and matrix notation the formal solution of
(\ref{a12a}) reads
\begin{equation} \label{a13}
\tilde{n} (\vec k,z) = \tilde{n}_0 (\vec k,z) +
\tilde{f}(\vec k,z),
\end{equation}
where the sure (deterministic) part is $
\tilde{n}_0 (\vec k,z) \equiv$ $ \Delta (\vec k,z)\, {\rm
e}^{i \vec k \cdot
\vec c + z}\, n(\vec k,0)$
and the fluctuating part is $
\tilde{f}(\vec k,z) \equiv$ $ \Delta (\vec k,z)\, \tilde{F}
(\vec k,z)$.
Here $ \Delta_{ij} (\vec k,z)$ and $ [\exp({i \vec k \cdot
\vec c})]_{ij}$ $ = \delta_{ij}
\exp({i \vec k \cdot \vec c_i})  $ are ma\-trices, and
$\tilde{n}
(\vec k,z)$,{ \em etc} are vectors with components
$\tilde{n}_i (\vec k, z), etc$. The matrix $\Delta$ is
defined as
\begin{equation} \label{a14}
 \Delta (\vec k,z) = [ \,{\rm e }^{i \vec k \cdot \vec c +
z} -1
 - \Omega]^{-1}.
\end{equation}
The sure part of (\ref{a13}) approaches the Chapman-Enskog
hydrodynamic solution of the Boltzmann equation for large
spatial and temporal scales. The fluctuating part of the
distribution function has correlations given by,
\begin{equation} \label{a15}
V^{-1} \langle \tilde{f}_i(\vec k,z)  \tilde{f}^\ast_j(\vec
k,z^\prime)
\rangle
=V^{-1}  \Delta_{i \ell} (\vec k,z) \langle
\tilde{F}_\ell(\vec k,z)
\tilde{F}^\ast_m(\vec k,z^\prime) \rangle  \Delta_{jm}^*
(\vec k,z^\prime),
\end{equation}
where the asterisk represent complex conjugation and where
$V$ is the number of nodes in the lattice. The correlation
function of the $\tilde{F}$'s is obtained by taking Fourier
and Laplace transforms of (\ref{a3}), and using
translational invariance with the result,
\begin{equation} \label{a17}
V^{-1} \langle \tilde{f}_i(\vec k,z) \tilde{f}^{\ast}_j(\vec
k,z^\prime)
\rangle
=  \Delta_{i \ell} (\vec k,z) B_{\ell m}  \Delta^\ast_{jm}
(\vec k,z^\prime)/
[\, 1 -  {\rm e} ^{-z -z^\prime}].
\end{equation}
\noindent
As we are interested in hydrodynamic space and time scales,
we take the limit $\vec k \to 0, z \to 0$. Then $ \Delta
(\vec k,z) \to -1/\Omega$, where the inverse of $\Omega$ is
only defined in the subspace orthogonal to the
zero-eigenfunctions, the so-called collisional invariants.
Next we invert the transforms in (\ref{a17}) to find the
fluctuations ${f}_i(\vec r,t) $ in the distribution function
as,
\begin{eqnarray} \label{a18}
 &\langle {f}_i(\vec r,t) {f}^\ast_j(\vec r\
^\prime,t^\prime) \rangle
 \simeq \delta(\vec r, \vec r\ ^\prime) \delta(t,t^\prime)
\Omega^{-1}_{i \ell} B_{\ell m} \Omega^{-1}_{jm}  =
&\nonumber \\
& - \delta(\vec r, \vec r\ ^\prime) \delta(t,t^\prime)
\left[ \left( \frac{1}{\Omega} + {1\over 2} \right)_{ij} g_j
+
\left( \frac{1}{\Omega} + {1\over 2} \right)_{ji} g_i
\right].&
\end{eqnarray}
The second equality is a consequence of (\ref{a5}). This is
the final result for the  fluctuations in the distribution
functions on hydrodynamic space and time scales. The
equation also shows that the term $\propto \Omega^2$ in
(\ref{j8}) and (\ref{a5}) directly transforms into the two
terms in (\ref{a18}) containing $1\over 2$, which constitute
the so called `propagation' part of the transport
coefficients.\\

\indent The result (\ref{a18}) enables us to calculate the
strength of the
 fluctuations in the dissipative  currents in the
fluctuating hydrodynamic equations, such as the stress
tensor or the heat current in lattice gases with energy
conservation. We illustrate this by calculating the
fluctuations in the shear stress  $\hat{\Pi}_{xy}(\vec r,t)
= \sum_i c_{xi} c_{yi} {f}_i(\vec r,t)$ with the help of
(\ref{a18}). This yields,
\begin{equation} \label{a19}
\langle \hat{\Pi}_{xy}(\vec r,t) \hat{\Pi}_{xy}(\vec
r\,^\prime,t^\prime)
\rangle =
2 {\langle c_x \! \mid \! c_x \rangle}
\, \nu \, \delta(\vec r, \vec r\ ^\prime)\,
\delta(t,t^\prime),
\end{equation}
where $\nu$ is the kinematic viscosity, given by
\begin{equation} \label{a20}
\nu = - {\langle c_x c_y \mid
\left( \frac{1}{\Omega} + {1\over 2} \right) \mid c_x c_y
\rangle}
 /{\langle c_x \! \mid \! c_x \rangle}.
\end{equation}
The brackets are defined as,
\begin{equation} \label{a21}
\nu = {\langle a \! \mid M \mid \! b \rangle} =  a_i\,
M_{ij}\, g_j\, b_j,
\end{equation}
which includes the definition of the inner product for
$M=1$.  This expression is the standard formula for the
kinematic viscosity as obtained from a lattice Boltzmann
equation [Brito et al 1991].  In lattice gas automata the
occupation numbers obey an exclusion principle, sothat $g_i
= \langle (\delta {n}_i)^2 \rangle_o = f^o_i( 1-f^o_i)$.  In
athermal lattice gases with or without rest particles the
single particle state distribution function $f^o_i =f$ is
independent of the speed $|\vec c_i|$ and equal to the
average density $\rho$ per node, divided by the number of
allowed velocity channels.

\section{Concluding Remarks}

\indent We offer several remarks to summarize and put the
above results
in context.

1) The noise considered here has been assumed to be white
and Gaussian, and the resulting Boltzmann-Langevin equations
define a discrete time Markov process. It is entirely
characterized once the deterministic dynamics (i.e., the
form of the Boltzmann collision operator $\bf V$) and the
noise intensity ${\bf B}$ is specified. Non-Gaussian noise
could be considered as well, and most of the results
obtained here for pair correlations still hold.  However,
higher order correlations would be different.

2) For stationary and quasi-stationary states the noise
intensity has been determined by the consistency between the
solution to the linearized Boltzmann-Langevin equation for
equal time pair correlations and a specified stationary
value for this correlation. The important new result here is
the fluctuation-dissipation theorem (\ref{j8}). It relates
the fluctuations to the matrix elements of the linear
collision operator and the equilibrium value of the
correlations $g(x,y)$, which are assumed to be given. The
relation is somewhat similar to that for continuous gases,
but there are some typical lattice effects, caused by the
discreteness of time and closely related to the so called
`propagation' viscosities of lattice gases.

3) The linear Boltzmann-Langevin equations imply associated
hydrodynamic equations with noise. The latter occurs as an
additional component to the stress tensor. The fluctuation
formula for the stress tensor (\ref{a19}), as derived from
the Langevin noise added to the lattice Boltzmann equation,
is in complete agreement with the results of Landau and
Lifschitz for the noise in fluid dynamics. Furthermore, the
explicit form of (\ref{a19}) identifies the noise intensity
in terms of the Boltzmann value of the transport
coefficient.

4) The examples at the end of section 2 concern models
satisfying the conditions of SDB, where the equilibrium
distributions are completely factorized and position and
velocity correlations are completely absent, as in
(\ref{a3b}). However, this is not the case in LGCA's
violating SDB, where computer simulations have shown
[H\'enon 1992,  Bussemaker and Ernst 1992]
 the existence of (strong) on-node velocity correlations and
(weaker) off-node spatial correlations in thermal
equilibrium. The existence of such correlations is not yet
understood. However, in driven diffusive systems [Zia et al
1993, Garrido et al 1990],
 where detailed balance is broken by imposing an external
bias field or by boundary conditions, the absence of
detailed balance does indeed give rise to long spatial
correlations.

5) As long as the analytic structure of the equilibrium
correlation function $g(x,y)$ is unknown, the
fluctuation-dissipation relation (\ref{j8}) is still valid
but does not provide any information about the noise
strength, $B(x,y)$. The same remarks apply when dealing with
stationary non-equilibrium states, supporting stationary
temperature gradients or shear rates. For real fluids the
equal time correlation functions in such states are long
ranged, and have been calculated from both Langevin and
kinetic theories [Schmitz 1988] for which the noise strength
is specified. As this information is not yet available for
LGCA's further exploitation of (\ref{j8}) is not possible.
Similar limitations apply to (\ref{j5}) that describes
fluctuations around non-stationary states. To calculate the
correlation functions it is necessary to know the noise
strength as a function of the non-equlibrium state, ${\bf
f}(t)$. Again, this information is available for real fluids
from kinetic theory but the corresponding analysis for
LGCA's is incomplete [Ernst and Dufty 1994].

6) So far we have discussed stable systems. For unstable
systems (phase separation, spinodal deocmposition, pattern
formation) the fundamental quantity is the equal time
correlation function $g(x,y,t)$ or its Fourier transform,
the time dependent structure factor $S(\vec k,t)$. In the
Cahn-Hilliard-Cook theory [Langer 1992]
 the Langevin equation has been used to study the onset of
instability, the wavelength of maximum growth, and initial
patterns. Such studies have been done for reals systems as
well as for LGCA's, quenched into a spatially uniform but
metastable or unstable state. For some LGCA's simulations
and analytic results (see Alexander et al [1992], Bussemaker
and Ernst [1993],  Ernst and Bussemaker {1994]) on $S(\vec
k,t)$ and $g(x,y,t)$ are becoming available, but studies on
fluctuations around such states have yet to be done. Once
$B(x,y\mid f(t))$ is understood on the basis of a
microscopic model, that information may be used for
mathematical modeling of $g(x,y,t)$ via the
fluctuation-dissipation relation (\ref{j5}).

\section*{Acknowledgements} The authors acknowledge
stimulating discussions
with H. J. Bussemaker.  One of us (M.E.) thanks the Physics
Department of the University of Florida, where this research
was carried out, for its hospitality in the summer of 1993,
and acknowledges support from a Nato Travel Grant for this
visit.

\end{document}